\documentclass[10pt,a4paper]{article}
\usepackage{bm, amssymb,pifont,cancel, amsmath,comment,color,authblk}
\usepackage[dvips]{graphicx}
\usepackage{multicol}
\newcommand{\cR}{{\cal R}}
\begin{document}
\begin{flushright}
WU-HEP-15-19
\end{flushright}
\begin{center}
{\Large{\bf{de Sitter vacuum from $R^2$ supergravity}}}\\ 
\vskip 7pt
\large{Fuminori Hasegawa}{\renewcommand{\thefootnote}{\fnsymbol{footnote}}\footnote[1]{E-mail address: corni\_in\_f@akane.waseda.jp}}, and \large{Yusuke Yamada}{\renewcommand{\thefootnote}{\fnsymbol{footnote}}\footnote[2]{E-mail address: yuusuke-yamada@asagi.waseda.jp}}\\
\vskip 4pt
{\small{\it Department of Physics, Waseda University,}}\\
{\small{\it Tokyo 169-8555, Japan}}
\end{center}
\begin{abstract}
We propose a supergravity model with a constrained curvature multiplet, which realizes the Starobinsky inflation and the de Sitter vacuum in the present universe simultaneously. Surprisingly, at the vacuum, the soft supersymmetry breaking scale for minimal supersymmetric standard model sector becomes the TeV scale, however, the gravitino mass scale becomes much higher than that of soft supersymmetry breaking. Such a hierarchical structure, which naturally avoids the gravitino problem, appears without introducing a new scale other than inflation scale. 
\end{abstract}
\begin{multicols}{2}
\section{Introduction}
The realization of de Sitter vacuum is one of the important issues in supergravity~(SUGRA), because the cosmological constant tends to be negative without supersymmetry (SUSY) breaking~\cite{Townsend:1977qa}. The positive cosmological constant, which is required to explain the accelerated universe we observe, can be realized with the SUSY breaking multiplets. It is known that SUSY breaking is simply realized with a nilpotent superfield $X$, which satisfies $X^2=0$ and leads to the nonlinearly realized SUSY~\cite{Volkov:1973ix}-\cite{Kuzenko:2010ef}. Such a nilpotent multiplet has received renewed interest in recent years from the viewpoint of inflation~\cite{Guth:1980zm}-\cite{Linde:1983gd} and de Sitter vacuum in SUGRA~\cite{Antoniadis:2014oya}-\cite{Ferrara:2015gta} and realization of de Sitter vacua in superstring (see~\cite{Kallosh:2014wsa,Kallosh:2015nia} and references therein). The component action of a constrained multiplet coupled to SUGRA has also been constructed in Refs.~\cite{Bergshoeff:2015tra}-\cite{Kallosh:2015tea}.

 Recently, it was found that the simplest de Sitter SUGRA model, in which a single nilpotent multiplet $X$ couples to SUGRA, is equivalent to the pure supergravity with a constrained curvature multiplet ${\cR}$~\cite{Dudas:2015eha,Antoniadis:2015ala}. In conformal SUGRA formalism~\cite{Kaku:1978nz}-\cite{supergravity}, the action is written as
 \begin{align}
 S=-3[S_0\bar{S}_0]_D+[S_0^3W_0]_F,\label{puredS}
 \end{align}
 with a constraint
 \begin{align}
 \left(\frac{\cR}{S_0}-\lambda\right)^2=0,\label{constraint}
 \end{align}
 where we have used the notation in Ref.~\cite{supergravity}, $[\cdots]_{D,F}$ are superconformal D- and F-term formulae respectively, $S_0$ is the chiral compensator multiplet, $\cR\equiv\frac{\Sigma(\bar{S}_0)}{S_0}$, where $\Sigma$ is a chiral projection operator, is a scalar curvature multiplet whose F-term contains the Ricci scalar $R$, $\lambda$ and $W_0$ are constants. With a specific choice of $W_0$ and $\lambda$, the de Sitter vacuum with a tunable cosmological constant is realized.
 
To realize the inflationary universe in such a simple de Sitter SUGRA model, we consider a simple extension of the model~(\ref{puredS}) by adding a higher curvature term~$[\frac{3}{\alpha}\cR \bar{\cR}]_D$ where $\alpha$ is a real parameter characterizing the inflation scale. Such a higher curvature extension is known as the Starobinsky model in old minimal SUGRA, which was first constructed in Ref.~\cite{Cecotti:1987sa} and developed in Refs.~\cite{Kallosh:2013xya}-\cite{Ketov:2013dfa}. The term leads to the higher curvature term $R^2$ where $R$ is the Ricci scalar. From the minimalistic viewpoint, we choose the parameter in Eq.~(\ref{constraint}) as $\lambda=\alpha$. As we will show in this letter, such a simple extension solves some problems in SUGRA inflation models and shows interesting features: 1.~{\it Inflation} is realized with the Starobinsky type potential, which predicts cosmological parameters consistent with the latest Planck results. 2.~The {\it de Sitter vacuum} is realized as in the original model. 3. The {\it estabilization problem} in the SUGRA Starobinsky model is absent due to the constraint. 4. A {\it TeV scale soft SUSY breaking} for minimal SUSY standard model (MSSM) sector is unexpectedly realized, however, 5. {\it extremely heavy gravitino} is simultaneously realized by which we can avoid the gravitino problem~\cite{Moroi:1993mb} which prevents the successful Big-Bang Nucleosynthesis after inflation.

We note that in the SUSY Starobinsky models, the SUSY breaking vacuum has been studied so far~\cite{Hindawi:1995qa}-\cite{Dalianis:2014aya}. However, the vacuum structures in such models are different from our model discussed below. The model in Ref.~\cite{Dalianis:2014aya} realizes the SUSY breaking vacuum with an unconstrained curvature multiplet. The resultant SUSY breaking scale in the model is the same as the inflation scale. Such a difference is originated from the constraint on $\cR$ we impose in our model.

The remaining parts of this letter are as follows. First, we show the higher curvature action with a constrained curvature multiplet, and also show the dual action which contains an inflaton multiplet $T$ and a constrained multiplet $S$. Then, we discuss the vacuum structure and inflation in Sec.~\ref{3}. After that, in Sec.~\ref{4}, we calculate the vacuum expectation values (VEVs) of F-terms, which are important for the spectra of soft SUSY breaking parameters for the MSSM sector. Finally, we conclude in Sec.~\ref{5}.
\section{Model}\label{2}
The action of the gravity sector is
\begin{align}
S=-3\left[S_0\bar{S}_0-\frac{1}{\alpha}\cR\bar{\cR}\right]_D+\left[S_0^3\left(W_0+L\left(\frac{\cR}{S_0}-\alpha\right)^2\right)\right]_F,\label{Model}
\end{align}
where $W_0$ is a constant and $L$ is a Lagrange multiplier chiral multiplet whose equation of motion leads to the constraint $(\frac{\cR}{S_0}-\alpha)^2=0$. We can rewrite the action~(\ref{Model}) by using a Lagrange multiplier $T$ as
\begin{align}
S=&-3\left[S_0\bar{S}_0\left(1-\frac{1}{\alpha}S\bar{S}\right)\right]_D+[S_0^3W_0]_F\nonumber\\
&+[S_0^3L(S-\alpha)^2]_F+\left[S_0^33T\left(S-\frac{\cR}{S_0}\right)\right].\label{dual1}
\end{align} 
The equation of motion of $T$ gives $S=\cR/S_0$, which reproduces the action~(\ref{Model}). By using the identity~\cite{Cecotti:1987sa}, $[TS_0\Sigma(\bar{S}_0)]_F=[S_0\bar{S}_0(T+\bar{T})]_D$, we can rewrite the action~(\ref{dual1}) as
\begin{align}
S=&-3\left[S_0\bar{S}_0\left(1+T+\bar{T}-\frac{1}{\alpha}S\bar{S}\right)\right]_D+[S_0^3(3TS+W_0)]_F\nonumber\\
&+[S_0^3L(S-\alpha)^2]_F.\label{dual2}
\end{align}
For simplicity of the following discussion, we rescale $S$ and $L$ as $S\to \sqrt{\frac{\alpha}{3}}S$ and $L\to \frac{3L}{\alpha}$, and define $M=\sqrt{3\alpha}$. Then the action~(\ref{dual2}) becomes
 \begin{align}
S=&-3\left[S_0\bar{S}_0\left(1+T+\bar{T}-\frac{1}{3}S\bar{S}\right)\right]_D\nonumber\\
&+[S_0^3(MTS+W_0)]_F+[S_0^3L(S-M)^2]_F.\label{dual3}
\end{align}
After superconformal gauge fixing with conventional conditions~\cite{Kugo:1982mr}, we find that this system is the standard SUGRA action with K\"ahler potential $K$ and superpotential $W$ given by 
\begin{align}
K=&-3\log\left(1+T+\bar{T}-\frac{|S|^2}{3}\right),\\
W=&MTS+W_0,
\end{align}
and $S$ satisfies a SUSY constraint
\begin{align}
(S-M)^2=0.\label{const}
\end{align}
 Here and hereafter, we use the Planck unit convention $M_{Pl}(\sim2.4\times 10^{18}{\rm GeV})=1$.
As we will show in Sec.~\ref{3}, the value of $W_0$ is set to be $W_0\sim M^2/6$ to realize the de Sitter vacuum with an almost vanishing cosmological constant. Then, we notice that the action contains {\it only two mass scales}, the Planck scale $M_{\rm Pl}=1$ and $M$ which corresponds to the inflaton mass scale as we will see below. 
\section{Inflation and vacuum structure}\label{3}
Let us discuss the structure of the scalar potential in the model~(\ref{dual3}) in this section. Before starting the discussion, we have to solve the constraint~(\ref{const}). This constraint is almost the same as one of the nilpotent multiplets, and leads to
\begin{align}
s=M+\frac{\psi^S\psi^S}{2F^S},\label{Sol}
\end{align}
where $s$, $\psi^S$, and $F^S$ are the scalar, the fermion, and the auxiliary field of $S$. Then, the scalar $s$ disappears in the physical system. Therefore, the instability problem of $s$ can be solved as in Ref.~\cite{Antoniadis:2014oya}.

Taking the condition~(\ref{Sol}) into account, we can calculate the scalar potential as
\begin{align}
V=&e^K\left(K^{I\bar{J}}D_IWD_{\bar{J}}\bar{W}-3|W|^2\right)|_{s=M}\nonumber\\
=&\frac{M^2}{(1+2{\rm Re}\tilde{T}-\frac{M^2}{3})^2}\left[\left|\tilde{T}-\frac{M^2}{3}\right|^2-\frac{M^4}{9}+\frac{M^2}{3}-2W_0\right],\label{potential}
\end{align}
where $\tilde{T}$ is the scalar component of $T$.

First, to realize vanishing cosmological constant, we choose $W_0$ as 
\begin{align}
W_0=\hat{W}_0\equiv\frac{M^2}{6}\left(1-\frac{M^2}{3}\right).\label{W0}
\end{align}
Then the scalar potential takes a simpler form:
\begin{align}
V=\frac{M^2}{(1+2{\rm Re}\tilde{T}-\frac{M^2}{3})^2}\left|\tilde{T}-\frac{M^2}{3}\right|^2.\label{potential2}
\end{align}
The potential minimum is located at $\tilde{T}=\frac{M^2}{3}$, and the cosmological constant vanishes at the point. Therefore, with $W_0=\hat{W}_0-\epsilon$ where $\epsilon\ll1$, the positive cosmological constant $\Lambda$ can be realized around $T\sim \frac{M^2}{3}$ with 
\begin{align}
\Lambda\sim \frac{2M^2}{(1+\frac{M^2}{3})^2}\epsilon.
\end{align}
Thus, we can confirm that the de Sitter vacuum with a small cosmological constant is realized in this model. In the following discussion, we neglect the cosmological constant because the observed value of $\Lambda$ is $\mathcal{O}(10^{-120})$ in Planck units.

Next, let us discuss the inflation. Neglecting $\Lambda$, the scalar potential becomes one in Eq.~(\ref{potential2}). We identify the inflaton as ${\rm Re}\tilde{T}\equiv t$, and ${\rm Im}\tilde{T}$ is stabilized at ${\rm Im}\tilde{T}=0$ during and after inflation. As we will see below, $M$ is set to be $M\sim \mathcal{O}(10^{-5})$, and then we can approximate the potential~(\ref{potential2}) as
\begin{align}
V=\frac{M^2t^2}{(1+2t)^2}.\label{apx}
\end{align}
The kinetic term of $\tilde{T}$ is given by $-K_{T\bar{T}}\partial_\mu \tilde{T}\partial^\mu\bar{\tilde{T}}=-\frac{3}{(1+2t)^2}(\partial_\mu t\partial^\mu t)+\cdots$ where the ellipses denote the kinetic term of the imaginary part of $\tilde{T}$. In terms of the canonically normalized inflaton $\phi=\sqrt{\frac{3}{2}}\log (1+2t)$, the scalar potential~(\ref{apx}) becomes
\begin{align}
V=\frac{M^2}{4}(1-e^{-\sqrt{\frac{2}{3}}\phi})^2,
\end{align}
which predicts the cosmological parameters consistent with the Planck 2015 results~\cite{Ade:2015lrj}. The observed amplitude of the scalar power spectrum requires that $M\sim 10^{-5}$.
\section{Spectra of soft SUSY breaking parameters and the gravitino problem}\label{4}
In this section, we discuss the spectra of soft SUSY breaking parameters in this model. The soft breaking parameters are determined by the VEVs of F-terms and the couplings between matter multiplets, $T$, and $S$. We can evaluate the VEVs of the F-terms of $T$ and $S$ as
\begin{align}
\langle |F^T|\rangle\equiv& \langle \sqrt{K_{T\bar{T}}F^T\bar{F}^{\bar{T}}}\rangle
=\frac{M^2}{2\sqrt{3}}\left[\frac{\sqrt{1-M^2}}{1+\frac{M^2}{3}}\right]\sim \frac{M^2}{2\sqrt{3}},\label{FT}\\
\langle |F^S|\rangle\equiv& \langle \sqrt{K_{S\bar{S}}F^S\bar{F}^{\bar{S}}}\rangle
=\frac{M^3}{\sqrt{3}}\sqrt{\frac{1+\frac{7M^2}{9}}{(1+\frac{M^2}{3})^3}}\sim \frac{M^3}{\sqrt{3}},\label{FS}
\end{align}
where we have used $F^I=-e^{\frac{K}{2}}K^{I\bar{J}}D_{\bar{J}}\bar{W}$, $\langle T\rangle=\frac{M^2}{3}$, $\langle S\rangle=M$, and $M\sim 10^{-5}\ll 1$. In the same way, the gravitino mass is evaluated as
\begin{align}
m_{3/2}=\langle e^{\frac{K}{2}}W \rangle=\frac{M^2}{6}\frac{1+\frac{5M^2}{3}}{\sqrt{(1+\frac{M^2}{3})^3}}\sim \frac{M^2}{6}.
\end{align}

To discuss possible matter couplings to $T$ and $S$, let us return to the original higher curvature action~(\ref{Model}). The matter coupled extension of the action~(\ref{Model}) is
\begin{align}
S=&-\left[S_0\bar{S}_0\tilde{\mathcal{N}}\left(\frac{\cR}{S_0},\frac{\bar{\cR}}{\bar{S}_0}, Q^I,\bar{Q}^{\bar{J}}\right)\right]_D\nonumber\\
&+\left[S_0^3\tilde{W}_m\left(\frac{\cR}{S_0},Q^I\right)+S_0^3L\left(\frac{\cR}{S_0}-\alpha\right)^2\right]_F\nonumber\\
&+\left[\frac{1}{4}\tilde{f}_{AB}\left(Q^I,\frac{\cR}{S_0}\right)\mathcal{W}^A\mathcal{W}^B\right],\label{matter1}
\end{align}
where $Q^I$ is a matter chiral multiplet, $\tilde{\mathcal{{N}}}$ is an arbitrary real function of $\frac{\cR}{S_0},Q^I$ and their conjugates, $\tilde{W}_m$ and $\tilde{f}_{AB}$ are holomorphic functions of the arguments. As in the procedure performed in Eqs.~(\ref{dual1}-\ref{dual3}), we can derive the dual action of that in Eq.~(\ref{matter1}) as
\begin{align}
S=&-3\left[S_0\bar{S}_0\left(T+\bar{T}-\frac{1}{3}\mathcal{N}(S,\bar{S}, Q^I,\bar{Q}^{\bar{J}})\right)\right]_D\nonumber\\
&+\left[S_0^3MTS+S_0^3W_m(S,Q^I)+S_0^3L(S-M)^2\right]_F\nonumber\\
&+\left[\frac{1}{4}f_{AB}(Q^I,S)\mathcal{W}^A\mathcal{W}^B\right],\label{matter2}
\end{align}
where $\mathcal{N}(S,\bar{S},Q^I,\bar{Q}^{\bar{J}})=\tilde{\mathcal{N}}(\frac{1}{3}MS,\frac{1}{3}M\bar{S},Q^I,\bar{Q}^{\bar{J}})$, $W_m(S,Q^I)=\tilde{W}_m(\frac{1}{3}MS,Q^I)$, and $f_{AB}(S,Q^I)=\tilde{f}_{AB}(\frac{1}{3}MS,Q^I)$. 

The most important feature is the form of the couplings between $T$ and $Q^I$. Those couplings in Eq.~(\ref{matter2}) take the so-called conformal sequestering form~\cite{Randall:1998uk,Giudice:1998xp}, with which the SUSY breaking effect by $T$ is never mediated to the matter sector $Q^I$. After superconformal gauge fixing, we find that the action~(\ref{matter2}) becomes the standard SUGRA action with the following K\"ahler and super-potential,
\begin{align}
K=&-3\log\left(T+\bar{T}-\frac{1}{3}\mathcal{N}\right),\\
W=&MTS+W_m,
\end{align}
and $S$ satisfies the constraint~(\ref{const}). For concreteness of the discussion, we assume the functions $\mathcal{N}$, $W_m$, and $f_{AB}$ as
\begin{align}
\mathcal{N}=&-3+|S|^2+\sum_{IJ}Q^I\bar{Q}^{\bar{J}}(\delta_I^J-c_I^{J}|S|^2),\label{N}\\
W_m=&\hat{W}_0+\sum_{IJK} \frac{1}{6}(y_{IJK}+\tilde{y}_{IJK}S)Q^IQ^JQ^K\nonumber\\
&+\sum_{IJ}\frac{1}{2}(\mu_{IJ}+\tilde{\mu}_{IJ}S)Q^IQ^J,\label{W}\\
f_{AB}=&\frac{\delta_{AB}}{g_A^2}(1-2h_AS),\label{f}
\end{align}
where $y_{IJK}$, $\tilde{y}_{IJK}$, $\mu_{IJ}$, $\tilde{\mu}_{IJ}$ and $h_A$ are complex valued constants, $c_{I\bar{J}}$ and $g_A$ are real constants, $\hat{W}_0$ is one in Eq.~(\ref{W0}), and $\delta_I^J$, $\delta_{AB}$ are Kronecker symbols. 

The soft SUSY breaking terms of the MSSM sector are given by 
\begin{align}
\mathcal{L}_{\rm soft}=&-\Biggl(\frac{1}{2}M_A \lambda^A\lambda^A+\frac{1}{6}a_{IJK}\tilde{Q}^I\tilde{Q}^J\tilde{Q}^K\nonumber\\
&+\frac{1}{2}B_{IJ}\tilde{Q}^I\tilde{Q}^J+{\rm h.c.}\Biggr)-m_{I\bar{J}}\tilde{Q}^I\bar{\tilde{Q}}^{\bar{J}},
\end{align}
where $\lambda^A$ is a gaugino, and $\tilde{Q}^I$ is a scalar component of $Q^I$. Let us consider the soft parameters in our case. By using the soft SUSY parameter formulae~\cite{Kaplunovsky:1993rd}, we obtain the following set of soft SUSY breaking parameters:
\begin{align}
M_A\sim&\langle |F^S|\rangle h_A,\label{GM}\\
a_{IJK}\sim&\langle |F^S|\rangle \tilde{y}_{IJK},\label{A}\\
b_{IJ}\sim&\langle |F^S|\rangle\tilde{\mu}_{IJ},\label{B}\\
m_{I\bar{J}}^2\sim&\langle |F^S|\rangle^2 c_{I\bar{J}}.\label{SM}
\end{align}
As we discussed below Eq.~(\ref{matter2}), the soft parameters do not depend on $F^T$ due to the conformal sequestering structure of the K\"ahler potential, which is always realized even if we choose different $\mathcal{N}$ and $W_m$.

Let us compare the scale of the soft SUSY breaking terms and that of gravitino mass. From Eqs. (\ref{GM})-(\ref{SM}), we find that the scale of soft SUSY breaking parameters are determined by $\langle |F^S|\rangle\sim M^3$, on the other hand, the gravitino mass scale is $m_{3/2}\sim M^2$ in Planck unit. As we discussed in Sec.~\ref{3}, the order of $M$ should be $\mathcal{O}(10^{-5})$ due to the normalization of the cosmic microwave background power spectrum. Therefore, surprisingly enough, the scales of the gravitino mass and the soft parameters are {\it predicted} as
\begin{align}
m_{\rm soft}\sim& M^3=\mathcal{O}(10^{-15})=\mathcal{O}(1)\ {\rm TeV},\\
m_{3/2}\sim&M^2=\mathcal{O}(10^{-10})=\mathcal{O}(10^5)\ {\rm TeV}.
\end{align}
Because of the large gravitino mass, one may think that the anomaly mediation~\cite{Randall:1998uk,Giudice:1998xp} dominates the soft SUSY parameters, however, we find that the anomaly mediation is much smaller than the $S$-mediated SUSY breaking by the following reason. It is known that the scale of anomaly mediation is characterized by the F-term of the compensator $F^{S_0}$~\cite{Bagger:1999rd}. In our case, the VEV of $F^{S_0}$ is evaluated as
\begin{align}
|\langle F^{S_0}\rangle|=\langle|\frac{1}{3}e^{\frac{K}{6}}(K_IF^I+3e^{\frac{K}{2}}\bar{W}) |\rangle=\frac{4M^4}{9(1+\frac{M^2}{3})^2},
\end{align}
which is much smaller than $\langle |F^S|\rangle\sim M^3$ and, therefore, we can neglect the anomaly mediated SUSY breaking effects. The suppression of $F^{S_0}$ is realized by the so-called no-scale structure of K\"ahler potential. Indeed, the largest F-term is one of $T$ as shown in Eqs.~(\ref{FT}) and (\ref{FS}).

Finally, we briefly discuss the reheating processes in this model. Unlike the case in Ref.~\cite{Terada:2014uia}, the inflaton $\tilde{T}$ dominates the SUSY breaking, and then we can regard the inflaton as the sgoldstino, which mostly decays into the pair of gravitinos $\psi_{3/2}$ with the decay rate~(see e.g.~Ref.~\cite{Jeong:2012en})
\begin{align}
\Gamma(\tilde{T}\to \psi_{3/2}+\psi_{3/2})\sim& \frac{e^K|D_TW|^2}{288\pi}\frac{m_{\rm inf}^5}{m_{3/2}^4}\sim \frac{M}{96\pi}.
\end{align}
After the reheating by the inflaton, the universe is dominated by the energy density of the gravitino. Such a scenario was studied in Ref.~\cite{Jeong:2012en} and called a {\it gravitino rich universe}. Next, the produced gravitinos decay into the MSSM sector with the decay rate,
\begin{align}
\Gamma(\psi_{3/2}\to {\rm MSSM}) \sim \frac{193}{384\pi}m_{3/2}^3.
\end{align}
The temperature at the gravitino decay is given by
\begin{align}
T_{3/2}=\left(\frac{\pi^2g_*}{90}\right)^{-\frac{1}{4}}\sqrt{\Gamma}\sim150{\rm GeV}\left(\frac{g_*}{80}\right)^{-\frac{1}{4}},\label{T}
\end{align}
where $g_*$ is the relativistic degrees of freedom at $T=T_{3/2}$. The decay temperature~(\ref{T}) is much higher than the one at the BBN and, therefore, the successful BBN can be realized in our setup. Further investigation of the cosmological history of this scenario will be shown in future work.    
\section{Summary}\label{5}
We have investigated a simple model which realizes {\it inflation} consistent with the Planck result, the {\it de Sitter vacuum} with a tunable cosmological constant, a {\it heavy gravitino} not leading to the problematic thermal history, and a {\it TeV scale soft SUSY breaking} of MSSM sector solving the little hierarchy problem.

We assumed a simple higher curvature action~(\ref{Model}) with a constraint on the scalar curvature multiplet $\cR$, which is equivalent to the one in Eq.~(\ref{dual3}). From the requirement of the almost vanishing cosmological constant, a parameter $W_0$ in Eq.~(\ref{dual3}) is fixed to the value~(\ref{W0}) as discussed in Sec.~\ref{3}. Therefore, our model contains {\it the only one dimensionful parameter} $M$, which is fixed to the value $M\sim \mathcal{O}(10^{-5})\sim \mathcal{O}(10^{13})\ {\rm GeV}$ by the Planck normalization of the scalar power spectrum. However, at the vacuum, two new scales appear; the gravitino mass scale $m_{3/2}\sim \mathcal{O}(10^5) {\rm TeV}$ and the soft SUSY breaking scale $m_{soft}\sim \mathcal{O}(1){\rm TeV}$. They originate from the nontrivial vacuum in our model as shown in Sec.~\ref{4}.

To confirm the model, the following issues remain. 1. We have to investigate more detailed thermal history of our scenario, which can be done by determining the couplings in Eqs.~(\ref{N})-(\ref{f}). We also have to construct the component action with the nontrivial condition~(\ref{Sol}). The action under such a nontrivial constraint was studied recently in Refs.~\cite{Bergshoeff:2015tra}-\cite{Kallosh:2015tea}. 2. To test the model with the forthcoming collider experiments, such as LHC Run2, the study of particle phenomenology is also required. 3. The origin of the constraint on $\cR$ is nontrivial, however, the understanding of it is important to realize our model in a UV theory.
\section*{Acknowledgements}
YY is supported by Research Fellowships for Young Scientists (No.26-4236), which is from the Japan Society for the Promotion of Science.

\end{multicols}

\begin{thebibliography}{99}
\bibitem{Townsend:1977qa} 
  P.~K.~Townsend,
  Phys.\ Rev.\ D {\bf 15}, 2802 (1977).
\bibitem{Volkov:1973ix} 
  D.~V.~Volkov and V.~P.~Akulov,
  Phys.\ Lett.\ B {\bf 46}, 109 (1973).
\bibitem{Deser:1977uq} 
  S.~Deser and B.~Zumino,
  Phys.\ Rev.\ Lett.\  {\bf 38}, 1433 (1977).
\bibitem{Ivanov:1978mx} 
  E.~A.~Ivanov and A.~A.~Kapustnikov,
  J.\ Phys.\ A {\bf 11}, 2375 (1978).
\bibitem{Rocek:1978nb} 
  M.~Rocek,
  Phys.\ Rev.\ Lett.\  {\bf 41}, 451 (1978).
\bibitem{Lindstrom:1979kq} 
  U.~Lindstrom and M.~Rocek,
  Phys.\ Rev.\ D {\bf 19}, 2300 (1979).
\bibitem{Komargodski:2009rz} 
  Z.~Komargodski and N.~Seiberg,
  JHEP {\bf 0909}, 066 (2009)
  [arXiv:0907.2441 [hep-th]]. 
\bibitem{Kuzenko:2010ef} 
  S.~M.~Kuzenko and S.~J.~Tyler,
  Phys.\ Lett.\ B {\bf 698}, 319 (2011)
  [arXiv:1009.3298 [hep-th]].
\bibitem{Guth:1980zm} 
  A.~H.~Guth,
  Phys.\ Rev.\ D {\bf 23}, 347 (1981).
\bibitem{Starobinsky:1980te} 
  A.~A.~Starobinsky,
  Phys.\ Lett.\ B {\bf 91}, 99 (1980)
\bibitem{Sato:1980yn} 
  K.~Sato,
  Mon.\ Not.\ Roy.\ Astron.\ Soc.\  {\bf 195}, 467 (1981).
\bibitem{Linde:1981mu} 
  A.~D.~Linde,
  Phys.\ Lett.\ B {\bf 108}, 389 (1982).
\bibitem{Albrecht:1982wi} 
  A.~Albrecht and P.~J.~Steinhardt,
  Phys.\ Rev.\ Lett.\  {\bf 48}, 1220 (1982).
\bibitem{Linde:1983gd} 
  A.~D.~Linde,
  Phys.\ Lett.\ B {\bf 129}, 177 (1983).
\bibitem{Antoniadis:2014oya} 
  I.~Antoniadis, E.~Dudas, S.~Ferrara and A.~Sagnotti,
  Phys.\ Lett.\ B {\bf 733}, 32 (2014)
  [arXiv:1403.3269 [hep-th]].
\bibitem{Ferrara:2014kva} 
  S.~Ferrara, R.~Kallosh and A.~Linde,
  JHEP {\bf 1410}, 143 (2014)
  [arXiv:1408.4096 [hep-th]].
\bibitem{Kallosh:2014via} 
  R.~Kallosh and A.~Linde,
  JCAP {\bf 1501}, no. 01, 025 (2015)
  [arXiv:1408.5950 [hep-th]].
\bibitem{Aoki:2014pna} 
  S.~Aoki and Y.~Yamada,
  Phys.\ Rev.\ D {\bf 90}, no. 12, 127701 (2014)
  [arXiv:1409.4183 [hep-th]].
\bibitem{Dall'Agata:2014oka} 
  G.~Dall'Agata and F.~Zwirner,
  JHEP {\bf 1412}, 172 (2014)
  [arXiv:1411.2605 [hep-th]].
\bibitem{Kallosh:2014hxa} 
  R.~Kallosh, A.~Linde and M.~Scalisi,
  JHEP {\bf 1503}, 111 (2015)
  [arXiv:1411.5671 [hep-th]].
  \bibitem{Scalisi:2015qga} 
  M.~Scalisi,
  arXiv:1506.01368 [hep-th].
\bibitem{Carrasco:2015pla} 
  J.~J.~M.~Carrasco, R.~Kallosh and A.~Linde,
  arXiv:1506.01708 [hep-th].
\bibitem{Ferrara:2015gta} 
  S.~Ferrara, M.~Porrati and A.~Sagnotti,
  Phys.\ Lett.\ B {\bf 749}, 589 (2015)
  [arXiv:1508.02939 [hep-th]].
\bibitem{Kallosh:2014wsa} 
  R.~Kallosh and T.~Wrase,
  JHEP {\bf 1412}, 117 (2014)
  [arXiv:1411.1121 [hep-th]].
\bibitem{Kallosh:2015nia} 
  R.~Kallosh, F.~Quevedo and A.~M.~Uranga,
  arXiv:1507.07556 [hep-th].
\bibitem{Bergshoeff:2015tra} 
  E.~A.~Bergshoeff, D.~Z.~Freedman, R.~Kallosh and A.~Van Proeyen,
  arXiv:1507.08264 [hep-th].
\bibitem{Hasegawa:2015bza} 
  F.~Hasegawa and Y.~Yamada,
  arXiv:1507.08619 [hep-th].
\bibitem{Kallosh:2015sea} 
  R.~Kallosh,
  arXiv:1509.02136 [hep-th].
\bibitem{Kallosh:2015tea} 
  R.~Kallosh and T.~Wrase,
  arXiv:1509.02137 [hep-th].
\bibitem{Dudas:2015eha} 
  E.~Dudas, S.~Ferrara, A.~Kehagias and A.~Sagnotti,
  arXiv:1507.07842 [hep-th].
\bibitem{Antoniadis:2015ala} 
  I.~Antoniadis and C.~Markou,
  arXiv:1508.06767 [hep-th].
  \bibitem{Kaku:1978nz} 
  M.~Kaku, P.~K.~Townsend and P.~van Nieuwenhuizen,
  Phys.\ Rev.\ D {\bf 17}, 3179 (1978).
\bibitem{Kaku:1978ea} 
  M.~Kaku and P.~K.~Townsend,
  Phys.\ Lett.\ B {\bf 76}, 54 (1978).
\bibitem{Townsend:1979ki} 
  P.~K.~Townsend and P.~van Nieuwenhuizen,
  Phys.\ Rev.\ D {\bf 19}, 3166 (1979).
\bibitem{Kugo:1982cu} 
  T.~Kugo and S.~Uehara,
  Nucl.\ Phys.\ B {\bf 226}, 49 (1983).
\bibitem{Kugo:1982mr} 
  T.~Kugo and S.~Uehara,
  Nucl.\ Phys.\ B {\bf 222}, 125 (1983).
\bibitem{Kugo:1983mv} 
  T.~Kugo and S.~Uehara,
  Prog.\ Theor.\ Phys.\  {\bf 73}, 235 (1985).
   \bibitem{supergravity}
 For a recent review, D. Z. Freedman and A. Van Proeyen. Supergravity. Cambridge University Press, 2012
\bibitem{Cecotti:1987sa} 
  S.~Cecotti,
  Phys.\ Lett.\ B {\bf 190}, 86 (1987).
\bibitem{Kallosh:2013xya} 
  R.~Kallosh and A.~Linde,
  JCAP {\bf 1306}, 028 (2013)
  [arXiv:1306.3214 [hep-th]].
\bibitem{Farakos:2013cqa} 
  F.~Farakos, A.~Kehagias and A.~Riotto,
  Nucl.\ Phys.\ B {\bf 876}, 187 (2013)
  [arXiv:1307.1137 [hep-th]].
\bibitem{Ketov:2013sfa} 
  S.~V.~Ketov,
  PTEP {\bf 2013}, 123B04 (2013)
  [arXiv:1309.0293 [hep-th]].
\bibitem{Ferrara:2013wka} 
  S.~Ferrara, R.~Kallosh and A.~Van Proeyen,
  JHEP {\bf 1311}, 134 (2013)
  [arXiv:1309.4052 [hep-th]].
\bibitem{Ketov:2013dfa} 
  S.~V.~Ketov and T.~Terada,
  JHEP {\bf 1312}, 040 (2013)
  [arXiv:1309.7494 [hep-th]].
\bibitem{Moroi:1993mb} 
  T.~Moroi, H.~Murayama and M.~Yamaguchi,
  Phys.\ Lett.\ B {\bf 303}, 289 (1993).
\bibitem{Hindawi:1995qa} 
  A.~Hindawi, B.~A.~Ovrut and D.~Waldram,
  Nucl.\ Phys.\ B {\bf 476}, 175 (1996)
  [hep-th/9511223].
\bibitem{Hindawi:1996qi} 
  A.~Hindawi, B.~A.~Ovrut and D.~Waldram,
  Phys.\ Lett.\ B {\bf 381}, 154 (1996)
  [hep-th/9602075].
\bibitem{Dalianis:2014aya} 
  I.~Dalianis, F.~Farakos, A.~Kehagias, A.~Riotto and R.~von Unge,
  JHEP {\bf 1501}, 043 (2015)
  [arXiv:1409.8299 [hep-th]].

\bibitem{Ade:2015lrj} 
  P.~A.~R.~Ade {\it et al.} [Planck Collaboration],
  arXiv:1502.02114 [astro-ph.CO].
\bibitem{Kaplunovsky:1993rd} 
  V.~S.~Kaplunovsky and J.~Louis,
  Phys.\ Lett.\ B {\bf 306}, 269 (1993)
  [hep-th/9303040].
\bibitem{Randall:1998uk} 
  L.~Randall and R.~Sundrum,
  Nucl.\ Phys.\ B {\bf 557}, 79 (1999)
  [hep-th/9810155].
\bibitem{Giudice:1998xp} 
  G.~F.~Giudice, M.~A.~Luty, H.~Murayama and R.~Rattazzi,
  JHEP {\bf 9812}, 027 (1998)
  [hep-ph/9810442].
\bibitem{Bagger:1999rd} 
  J.~A.~Bagger, T.~Moroi and E.~Poppitz,
  JHEP {\bf 0004}, 009 (2000)
  [hep-th/9911029].
\bibitem{Terada:2014uia} 
  T.~Terada, Y.~Watanabe, Y.~Yamada and J.~Yokoyama,
  JHEP {\bf 1502}, 105 (2015)
  [arXiv:1411.6746 [hep-ph]].
\bibitem{Jeong:2012en} 
  K.~S.~Jeong and F.~Takahashi,
  JHEP {\bf 1301}, 173 (2013)
  [arXiv:1210.4077 [hep-ph]].
\end{thebibliography}
\end{document}